# A Review and Outlook of Energy Consumption Estimation Models for Electric Vehicles

Yuche Chen[1*], Guoyuan Wu[2], Ruixiao Sun[1], Abhishek Dubey[3], Aron Laszka[4], Philip Pugliese[5]
[1]Department of Civil and Environmental Engineering, University of South Carolina, Columbia, SC 29209
[2]Department of Electrical and Computer Engineering, University of California, Riverside, CA 92507
[3]Department of Electrical and Computer Engineering, Vanderbilt University, Nashville, TN
[4]Department of Computer Science, University of Houston, Houston, TX
[5]Chattanooga Area Regional Transportation Authority, 1617 Wilcox Blvd, Chattanooga, TN 37406
* Corresponding author: Phone: +1 (803) 777-9105; e-mail: chenyuc@cec.sc.edu

## Abstract

Electric vehicles (EVs) are critical to the transition to a low-carbon transportation system. The successful adoption of EVs heavily depends on energy consumption models that can accurately and reliably estimate electricity consumption. This paper reviews the state of the art of EV energy consumption models, aiming to provide guidance for future development of EV applications. We summarize influential variables of EV energy consumption in four categories: vehicle component, vehicle dynamics, traffic, and environment-related factors. We classify and discuss EV energy consumption models in terms of modeling scale (microscopic vs. macroscopic) and methodology (data-driven vs. rule-based). Our review shows trends of increasing macroscopic models that can be used to estimate trip-level EV energy consumption and increasing data-driven models that utilize machine learning technologies to estimate EV energy consumption based on a large volume of real-world data. We identify research gaps for EV energy consumption models, including the development of energy estimation models for modes other than personal vehicles (e.g., electric buses, trucks, and nonroad vehicles), energy estimation models that are suitable for applications related to vehicle-to-grid integration, and multiscale energy estimation models as a holistic modeling approach.

**Key Words:** electric vehicles, energy consumption estimation, machine learning model

## 1. Introduction

The transportation sector is a major energy consumer and contributor to air pollution. Governments around the world are taking steps to address energy and air pollution problems caused by transportation. A portfolio of strategies should be employed to mitigate transportation-related air pollution and reduce transportation's dependence on fossil fuels [1]. Transportation electrification is among the approaches promoted by industry, public agencies, and research communities. Electric vehicles (EVs) are considered one option to achieve low-carbon transportation systems. Countries around the globe are setting aggressive targets to promote EVs or even proposing to ban the future sale of internal combustion engine vehicles [2]. Norway, for example, wants EVs to account for 100% of its new-car sales by 2025. China aims to reach 7 million annual EV sales by 2025, which is equivalent to one fifth of its domestic market demand. France, United Kingdom, and California in the United States have proclaimed that they will end sales of internal combustion engine vehicles by 2040. The automotive industry expects EVs to become the major powertrain in the vehicle market by 2030 [3].

Despite the environmental benefits and rapid growth of EVs in the global market, "range anxiety" (i.e., user's concern about the insufficient all-electric range of an EV to reach a destination or charging point) is considered a major barrier that limits their wide adoption [4]. Reliable and accurate estimation of EV energy consumption can significantly mitigate range anxiety and help EV users can arrange their itineraries accordingly [5]. Additionally, vehicle-to-grid integration has drawn a tremendous amount of attention from research and industry communities in recent years, wherein EVs can communicate with the power grid to provide short-term demand response services that can balance loads (e.g., peak shaving) in the grid [6]. In this case, the EV energy consumption model can play an important role because it allows optimal management of EV battery charging and discharging activities with the consideration of integrated system efficiency in terms of energy use and transportation needs.

However, estimating the electricity consumption of an EV is a challenging problem. Various factors influence electricity consumption. In addition, the model setup can significantly vary depending on the granularity (in both time and space) of estimation. Over the years, a considerable amount of research has been conducted to gather insights into energy consumption estimation modeling of EVs. The majority of models can be classified from two perspectives: modeling scale and modeling methodology. Modeling scale refers to the spatial–temporal resolution of energy estimation results, which can be as detailed as an energy consumption rate (e.g., kWh per second) or averaged at the individual road link or trip level (e.g., kWh per mile). In literature, modeling scale is determined based on both the purpose of study and data availability. Methodologies of existing models can be roughly classified into rule-based and data-driven. Rule-based models adopt a "white-box" approach that follows some fundamental physics laws and mimics the dynamics and interactions of various vehicle and powertrain components to estimate energy consumption. Data-driven models draw on a "black-box" approach so that users do not need to understand the physical process of electricity generation and consumption, or even the principles governing vehicle dynamics and powertrain operation, but instead rely on the exploration of the statistical relationship between inputs and energy outputs with certain assumptions or statistical techniques.

This study aimed to provide a broad perspective of EV electricity consumption estimation and support the improvement of models and development of emerging EV applications. This review focused on papers published after 2010 because this decade witnessed a surge of publications on EVs. The remainder of the paper is organized as follows. Section 2 provides background information on mechanics and components of EVs. Section 3 involves a taxonomy analysis of influential variables for EV energy estimation. Section 4 and Section 5 classify and discuss existing EV electricity consumption models in the literature based on their modeling scale and modeling methodology. Section 6 and Sections 7 summarize possible applications of EV electricity consumption models and offer concluding thoughts.

## 2. Background

### *2.1. EV Types and Configuration*

In a broad definition, EVs refer to road vehicles whose propulsion involves electricity [7], including battery EVs (BEVs); hybrid EVs; plug-in hybrid EVs; and fuel-cell EVs. Figure 1 illustrates a general EV configuration, which is composed of three major subsystems: electric propulsion, energy source, and auxiliary. As shown in the figure, the electric propulsion subsystem consists of motor(s), transmission, power converter, and electronic control units. The energy storage unit, energy management unit, and energy refueling unit comprise the energy source subsystem. In practice, the most widely adopted energy storage device for EVs is a battery, due to their characteristics in terms of high energy density, compact size, and reliability [8]. Other devices may include an ultra-capacitor, flywheel, and hydrogen tank, which can be utilized as an auxiliary energy source or hybrid energy source [9,10]. The auxiliary subsystem involves the auxiliary power supply unit, power steering unit, and A/C control unit.

In this study, we focused on BEVs, which solely rely on energy stored in battery packs to provide power to the drivetrain. Therefore, their range depends directly on the battery capacity and other factors, including vehicle characteristics (e.g., configuration, weight), driving style, roadway conditions, and weather. To date, there are BEV configurations that may vary in: (a) the number of motors used (i.e., single motor, dual motors, and four motors); (b) the motor–transmission connection, such as multigear transmission with clutch, fixed gearing with or without differential, and in-wheel motors; and (c) the position of tractive power provision, i.e., front-wheel drive, rear-wheel drive, and all-wheel drive. The selection of different configurations mainly depends on the consideration of size, compactness, weight, cost, reliability, and performance (e.g., maximum cruising speed, gradeability, and acceleration) [11].

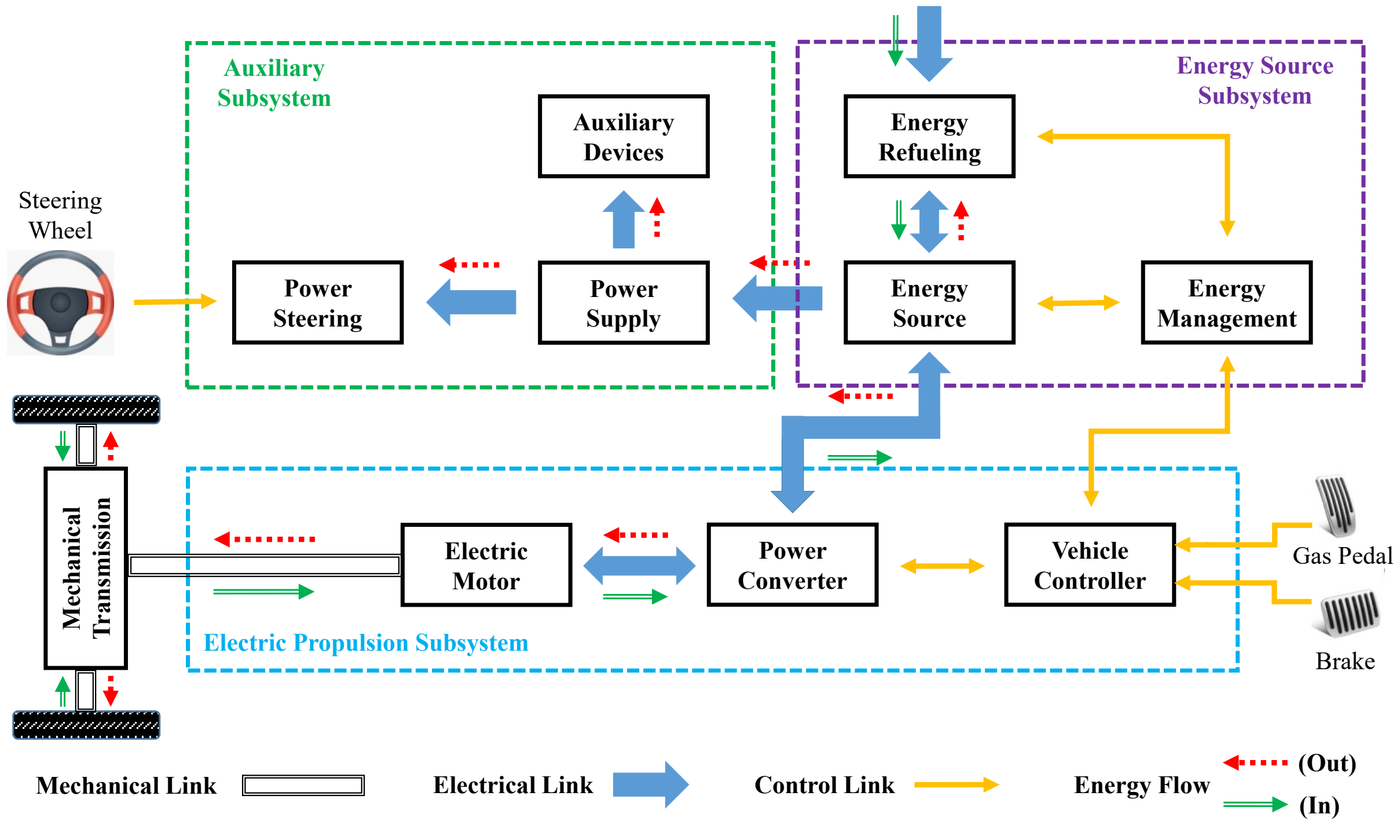

**Figure 1.** A representative configuration of EV.

*2.2. BEV Energy Consumption and Regenerative Braking*

Due to limited charging facilities and battery capacities, as well as long charging times, estimating BEVs' energy consumption is critical for not only environmental sustainability but also market adoption. For BEVs, energy consumption is an integration of the power output measured at the battery terminals (unit in kWh), where the battery charging and discharging modes should be handled separately [12].

During the propulsion mode, the batter power output $P_{out}$ (W) can be estimated by dividing the tractive power at wheels $P_{wheel}$ (W) by the powertrain's efficiency, which may consider the power losses in motor drive and transmission. The tractive power at the wheels is the product of the vehicle speed $v$ (m/s) and tractive force at the wheels $F_{wheel}$ (N), which can be approximated as the sum of the rolling resistance $F_{rr}$ (N), the aerodynamic force $F_{ad}$ (N), the road gradient force $F_{rg}$ (N), and the acceleration force $F_{accel}$ (N). More specifically,

$$P_{out} = \frac{P_{wheel}}{\eta_m \eta_t} = \frac{v \cdot F_{wheel}}{\eta_m \eta_t} = \frac{v \cdot \left(F_{rr} + F_{ad} + F_{rg} + F_{accel}\right)}{\eta_m \eta_t}$$

$$= \frac{v}{\eta_m \eta_t}\left(C_r mg\cos\alpha + \frac{\rho_a}{2} C_d A_f v^2 + mg\sin\alpha + m\delta \frac{dv}{dt}\right) \quad (1)$$

where $\eta_m$ and $\eta_t$ represent the efficiencies of electric motor(s) and transmission, respectively; $C_r$ and $C_d$ are the coefficients associated with rolling resistance and aerodynamic drag, respectively; $m$ (kg) is the vehicle mass; $g$ (9.81 m/s$^2$) is the standard gravitational acceleration; $\alpha$ (rad) stands for the road gradient; $\rho_a$ (1.2 kg/m$^3$) refers to the air density; $A_f$ (m$^2$) is the vehicle's effective cross-sectional area; and $\delta$ (constant) is the vehicle rotational inertial factor. Note that the nontraction load (e.g., A/C or lighting load) is not considered in the equation, and wheel slip is ignored.

The battery charging mode for BEVs usually occurs during coasting and braking, when the vehicle's kinetic energy—wasted in conventional vehicles—can be partially utilized to generate electricity to the supply side, which is known as regenerative braking. In other words, part of the braking energy can be recovered by operating the motor as a generator and transferred power into the battery. During the regenerative braking mode, the power at the battery terminals can be estimated as

$$P_{in} = \frac{kv}{\eta_m \eta_t}\left(C_r mgcos\,\alpha\ + \frac{\rho_a}{2} C_d A_f v^2 + mgsin\,\alpha\ + m\delta \frac{dv}{dt}\right) \qquad (2)$$

where $k$ ($0 < k < 1$) is the regenerative braking factor, indicating the percentage of the overall braking energy that can be recovered by the electric motor(s). In BEVs, the regenerative braking system has to couple with the friction brake, because the regenerative braking itself is not capable of generating enough power to fully stop the vehicle and the addition of friction brake can serve a safety purpose. Therefore, the total energy consumption from the battery $E_{batt}$ related to the motion of BEVs is

$$E_{batt} = \int_t^T P_{batt}\, dt \qquad (3)$$

Where $P_{batt} = P_{out}$ in traction mode and $P_{batt} = P_{in}$ in braking mode.

## 3. Taxonomy of Influential Variables on EV Energy Estimation

### *3.1. Types of Variables*

The energy consumption of an electric vehicle is influenced by a wide range of variables, which can be classified into four categories: (a) vehicle component related, (b) vehicle dynamics related, (c) traffic related, and (d) environment related. These variables are used in both disaggregated (i.e., original or instantaneously measured) and statistically or spatiotemporally aggregated formats.

#### 3.1.1. Vehicle Component-Related Variables

Vehicle component-related parameters govern the operating states of key parts for propulsion (e.g., electric motors, mechanical transmissions) and energy flows in the energy storage and auxiliary subsystems (Figure 1). For example, motor and transmission efficiencies determine the portion of generated energy from the source that can be used for propulsion [13-16, 84, 85, 95, 100, 105]. They vary based on specific configurations of EVs and motor and transmission technology. The battery state of charge (SOC) is found to affect the energy consumption rate of EVs [13, 19-22]. Studies have shown that the battery SOC can influence the instantaneous battery charging and discharging mechanism and efficiency; thus, it is also considered a critical explanatory variable [13, 22]. Other studies found that the initial battery SOC can aggravate or mitigate the range anxiety of EV drivers and subsequently adjust their driving behaviors, which can affect energy consumption [19-21]. Battery quality, i.e., degradation, is used to estimate changes in the trip-level energy consumption rate for EVs at different ages [103]. The auxiliary power that supports the operation of air conditioning, radio, monitor panel, and lights is nontrivial under certain environmental conditions. Studies have assumed auxiliary load to be either at constant load or estimated using environmental conditions based on real-time measured auxiliary load data [14, 17, 20, 23, 112, 113]. With the advent of enabling technologies for connected and automated vehicles, such as on-board sensors (e.g., LiDAR, camera), edge computing and processing units (e.g., GPU), and wireless communication devices, the energy consumption on these auxiliary loads may raise additional concerns. This topic could attract further attention by researchers and engineers in the future. In addition, studies have directly built statistical relationships between vehicle specifications (e.g., engine size, engine technologies, transmission type and efficiency) and energy consumption [83]. Coefficients of rolling resistance and aerodynamics are included in models that estimate EV energy at each second of driving according to the laws of physics, as shown in Equation (1) [86, 91-94, 104-106].

3.1.2. Vehicle Dynamics-Related Variables

Vehicle dynamics cover factors that reflect the motion (including speed, acceleration, and tractive or brake torque) of a vehicle or flow of vehicles. The laws of physics govern direct relationship between these factors and (kinetic) energy demanded by vehicles. Therefore, these variables are commonly used in EV energy estimation models. In existing literature, vehicle dynamics data are seen at an instantaneous level (e.g., every second) or certain aggregated levels (e.g., trip, road link, 5 minutes, etc.). Speed serves as a key parameter to estimate the road loads that are physically related to rolling resistance, aerodynamic drag, and road gradient as depicted in Equation (1) [13-16,18]. Instantaneous speed and its higher orders (up to the third order) show a strong correlation with instantaneous EV energy consumption [22, 24-27, 84, 85, 93-95, 100, 104-106]. When energy consumption is estimated at the trip level, average speed [19, 28, 29, 101] and its higher order [30-34] are considered. Other statistics related to speed and acceleration have also been used by researchers for EV energy consumption estimation. For example, studies [35, 36] noted the highest instantaneous speed and acceleration as surrogate driving behavior modes of drivers and used those features to estimate trip-level energy consumption of EVs. Distribution of speed during a trip is considered a metric to represent the driving behavior of drivers as a way to estimate energy consumption of EVs [87]. Profile of speed trajectory is used to estimate possible regenerative braking potential of EVs [88]. Kinetic energy and its change are also highly related to the energy consumed by EVs in motion. Qi et al. [37] used cumulative positive or negative changes in the kinetic energy rate (as described below) during a trip as influential variables to estimate EV energy consumption and achieved reasonable estimation performance.

$$PKE = \frac{\sum_{i=1}^{N-1} max\,(v_{i+1}^2 - v_i^2, 0)}{\sum_{i=1}^{N-1}(d_{i+1} - d_i)} \tag{4}$$

and

$$NKE = \frac{\sum_{i=1}^{N-1} min\,(v_{i+1}^2 - v_i^2, 0)}{\sum_{i=1}^{N-1}(d_{i+1} - d_i)} \tag{5}$$

where $d_i$ is the cumulative travel distance up to the $i$-th step. Vehicle-specific power is another conventionally defined term to represent the instantaneous vehicle tractive power normalized by the mass. Studies either try to establish relationships between instantaneous vehicle-specific power and energy consumption or distribution of vehicle-specific power over a short driving period, also referred as snippets, and average energy consumption rate in that period [33,38].

### 3.1.3. Traffic Conditions-Related Variables

Traffic conditions, such as downstream traffic signal status, congestion levels, and vehicle type mix in traffic flow, can influence EV energy consumption. Particularly, they are used to estimate or validate vehicle dynamics along the downstream segment or rest of the travel route, thus improving overall energy consumption estimation. Traffic conditions-related factors can be classified as categorical and interval variables. Categorical variables determine whether a trip is conducted in certain time or spatial resolution, e.g., time of day (i.e., peak hours vs. nonpeak hours), day of week (such as weekdays, weekends, or holidays), or month (e.g., seasonal effect). Fetene et al. [19] built a multiple linear regression model that includes “rush hour” as a dummy variable to identify whether a trip happened during peak periods (in the morning or afternoon). Masikos et al. [20] proposed a general regression neural network [39] model and used categorical variables to represent trip time by day of the week, month of the year, and hour of the day, and the model results showed the statistical significance of those variables in estimating EV energy consumption. Interval variables represent traffic conditions as a function of continuous vehicle dynamics or overall traffic states. The ratio of idle time or number of stops over travel time can be used as an indicator of traffic conditions (more congested if the ratio is higher) during a trip. Studies [28, 35, 40, 102] used this variable and found it to be statistically significant in EV energy models. Efforts have been made to create indexes of congestion that can be used to estimate energy consumption of EVs. Other studies [28, 89, 99] defined a congestion index (i.e., mean vehicle speed divided by standard deviation of speed) and found its significance in the model.

#### 3.1.4. Environment-Related Variables

Environment-related factors represent information about roadway characteristics or meteorological conditions. These variables influence energy consumption by introducing disturbance to the road or auxiliary loads (e.g., A/C power) for EVs. Widely used variables include road grade [84, 85], road type [19, 22, 41], wind direction, wind speed [23, 19, 41], ambient temperature and humidity [20, 21], and lighting conditions [42]. For example, changes in road elevation (i.e., road grade) affect tractive forces needed to overcome road gradient resistance. With advancement in outdoor positioning technology, road grade information becomes available in real time, and a plethora of studies have used it in EV energy estimation models at either the second-by-second level [15, 16, 17] or trip level [20, 23, 27, 28, 31, 32, 34, 40, 41, 43, 44, 84, 85, 92, 100, 101,106]. Another commonly used roadway characteristic-related variable in existing studies is road type, i.e., freeway vs. arterial [19, 22, 41, 90]. Infrastructure attributes of roads, such as traffic lights and speed limits, are used as continuous independent variables to estimate energy consumption of EVs traveling on roads [98]. Temperature and humidity are meteorological variables that may affect auxiliary power for heating or cooling demand and the operational performance of battery packs in EVs. Because meteorological conditions change gradually over a relatively long time, related variables are usually included in trip-level energy consumption estimation models [19, 20, 23, 27-29, 31, 41, 45, 96, 97], with the focus of their impacts on auxiliary power demand. Sun et al. tried to explore the relationship between meteorological parameters and battery performance by measuring the temperature at battery cells, but they did not include these variables in their proposed model [21]. In a regression model developed by Liu et al. [42], a dummy variable was adopted to represent day or night, and the lighting condition had a strong correlation with EV energy consumption. Ambient temperature and humidity are either measured or estimated based on longitude and latitude of driving location to estimate potential energy consumption for in-cabin cooling and heating [96, 97, 107].

### *3.2. Aggregation and Disaggregation of Variables*

Influential variables for EV energy estimation models can be also classified into two categories, i.e., disaggregated data and aggregated data. Disaggregated variables refer to data that have the same time or spatial interval as in the collection experiment. The frequency of data collection varies depending on the nature of studies. Data collected at 1 second or finer intervals are commonly seen in studies that collect data using on-board diagnostics equipment. Common disaggregated input data include 1Hz vehicle speed and acceleration [22, 24, 25, 27, 30], 1Hz vehicle specific power [26, 54], 1Hz kinetic energy [37], 1Hz road grade [15, 16], 1Hz battery state of charge [13, 22], or certain other statistics (e.g., maximum or minimum speed) of these data [19]. Other variables are usually collected at a large time interval because their values do not change frequently, which include rush house index (whether the travel time is during rush hour) [19,42]; congestion index or ratio of idling or stops during a trip or a link travel [28]; road type [19,22]; meteorology conditions of wind, humidity, and temperature [20, 21, 23, 27-29, 31]; infrastructure attributes (such as whether a road has traffic lights) [41]; and vehicle attributes (such as weight) [20].

**Table 1. Summary of Literature on EV Energy Prediction for Microscale Models (red) and Macroscale Models (blue)**

| Year | Veh. Type | Methodology | | Predicting Variable | | | | Data Source | Ref. |
|---|---|---|---|---|---|---|---|---|---|
| | | Data-driven | Rule-based | Dynamics | Traffic | Environment | Component | | |
| 2011 | PC* | X | | X | | X | X | Simulation | [85] |
| 2011 | PC | X | | X | | | | Simulation | [88] |
| 2011 | PC | | X | X | | | X | Simulation | [93] |
| 2012 | PC | X | | X | | | | Real-world | [87] |
| 2012 | PC | X | | X | | X | X | Real-world | [101] |
| 2012 | PC | | X | X | | | X | Simulation | [104] |
| 2013 | PC | X | | X | | X | | Real-world | [90] |
| 2013 | PC | | X | X | | | X | Simulation | [91] |
| 2013 | PC | X | | | X | | | Simulation | [99] |
| 2014 | PC | X | | X | | | X | Simulation | [86] |
| 2014 | PC | X | | X | | | | Real-world | [26] |
| 2014 | PC | X | | X | | | | Real-world | [43] |
| 2014 | PC | | X | X | | | X | Simulation | [46] |
| 2014 | PC | | X | X | | | X | Simulation | [94] |
| 2014 | PC | X | | X | X | X | X | Real-world | [108] |
| 2015 | PC | | X | X | | X | X | Real-world | [92] |
| 2015 | PC | | X | X | | | X | Simulation | [14] |
| 2015 | PC | | X | X | | | X | Real-world | [16] |
| 2015 | PC | X | | X | X | X | X | Real-world | [20] |
| 2015 | PC | X | | X | | | | Real-world | [22] |
| 2015 | PC | X | | X | | | | Real-world | [25] |
| 2015 | PC | X | | X | X | | X | Real-world | [40] |
| 2015 | PC | | X | X | | X | X | Simulation | [84] |
| 2015 | PC | X | | | X | X | | Real-world | [89] |
| 2015 | PC | | X | X | | | X | Simulation | [95] |
| 2015 | PC | X | | | | X | | Real-world | [96] |
| 2015 | PC | X | | | | X | | Real-world | [98] |
| 2015 | PC | | X | X | | | X | Real-world | [101] |
| 2016 | PC | | X | X | | | X | Simulation | [13] |
| 2016 | PC | | X | X | | X | X | Real-world | [15] |
| 2016 | PC | X | | X | X | X | X | Real-world | [28] |
| 2016 | PC | X | | X | X | X | X | Real-world | [42] |
| 2016 | PC | | X | X | | | X | Real-world | [48] |
| 2016 | PC | X | | X | | | | Real-world | [54] |
| 2016 | PC | X | | X | | | | Real-world | [56] |
| 2016 | PC | | X | X | | | X | Simulation | [105] |
| 2017 | PC | | | X | X | X | X | Real-world | [19] |
| 2017 | PC | X | | X | | X | X | Real-world | [24] |
| 2017 | PC | X | | X | | | X | Real-world | [27] |
| 2017 | PC | X | | X | | | X | Real-world | [30] |
| 2017 | PC | X | | X | | X | X | Simulation | [34] |
| 2017 | PC | X | | X | | X | | Real-world | [44] |
| 2017 | PC | X | | X | | X | | Real-world | [45] |
| 2017 | PC | X | | X | | | | Real-world | [53] |
| 2017 | PC | X | | X | | | X | Real-world | [55] |
| 2017 | Train | | X | X | | | X | Real-world | [60] |
| 2017 | PC | X | | X | | X | | Real-world | [107] |
| 2018 | PC | X | | X | | X | | Real-world | [17] |
| 2018 | PC | X | | X | | | X | Real-world | [29] |
| 2018 | PC | X | | X | | X | X | Real-world | [31] |
| 2018 | PC | X | | X | X | | | Real-world | [35] |
| 2018 | Bus | X | | X | | | | Real-world | [36] |
| 2018 | PC | X | | X | | X | | Real-world | [37] |
| 2018 | PC | | X | X | | | X | Real-world | [41] |
| 2018 | PC | X | | X | | | | Simulation | [47] |
| 2018 | PC | | X | X | | | X | Simulation | [50] |
| 2018 | PC | X | | X | | X | | Real-world | [57] |
| 2018 | Truck | | X | X | | X | X | Real-world | [106] |
| 2019 | Non-Road | X | | X | X | | | Real-world | [18] |
| 2019 | PC | X | | X | | | X | Real-world | [21] |
| 2019 | Bus | X | | X | X | | X | Simulation | [23] |
| 2019 | PC | | X | X | | | X | Real-world | [32] |
| 2019 | PC | X | | X | | | | Real-world | [33] |
| 2019 | PC | | X | X | | | X | Simulation | [49] |
| 2019 | PC | X | | X | | | | Real-world | [51] |
| 2019 | PC | X | | X | | X | | Real-world | [52] |
| 2019 | PC | X | | X | | | X | Real-world | [58] |
| 2019 | Truck | | X | X | | X | X | Simulation | [59] |
| 2019 | PC | | X | X | | X | | Real-world | [61] |
| 2019 | Bus | | X | X | | X | X | Real-world | [62] |
| 2019 | PC | | X | X | | | X | Simulation | [63] |
| 2019 | PC | | X | X | | | X | Real-world | [64] |
| 2019 | PC | X | | X | | X | X | Simulation | [65] |
| 2019 | PC | X | | X | | X | X | Simulation | [66] |
| 2019 | PC | X | | X | | | X | Real-world | [67] |
| 2019 | PC | X | | | | X | | Real-world | [97] |
| 2019 | PC | X | | | | | X | **Simulation** | **[83]** |
| 2019 | PC | X | | X | | X | X | **Real-world** | **[100]** |
| 2019 | PC | X | | X | X | | X | **Real-world** | **[102]** |
| 2019 | PC | X | | | | | X | **Real-world** | **[103]** |
| 2019 | PC | X | | X | X | X | | **Simulation** | **[109]** |

Note: PC = passenger car.

Aggregated variables refer to data that are aggregated after initially collected and presented at a different time interval. With on-board diagnostics or GPS technologies, speed and acceleration data are easily obtained at 1Hz level and then averaged to different minute or hour levels [21, 23, 28, 29, 31, 34-36, 40, 41, 43]. In addition, state and local transportation authorities routinely collect and publish traffic flow and volume information of roads in their jurisdictions, which are usually aggregated at every 5, 15, or 30 minutes or 1 hour.

## 4. Modeling Scale

EV energy estimation models can be classified based on their modeling scale, which further determine areas of application.

Microscopic-scale models can estimate energy consumption of EV at high frequency, typically 1Hz. Thus, they have been widely used in applications related to microscopic vehicle dynamics and optimal control of EV strings or traffic operations involved with EVs. A representative example is EV eco-driving, which uses microscopic models to optimize real-time vehicle control, particularly in congestion mitigation on corridors or signalized intersections [24, 47, 56, 109]. EV routing studies used microscopic energy estimation models to dynamically determine energy efficient routes [13, 59, 108]. Microscale models were developed for evaluating energy implications of EV in traffic simulation [15, 26, 30, 32].

Macroscopic models can explore the relationship between energy consumption and characteristics of driving at an aggregated spatial or temporal span. Thus, they are used in applications that require energy information of EVs in similar spatial and temporal space, such as EV fleet management, region-wide planning of charging infrastructure or EV adoption, large-scale EV-related energy portfolio prediction, etc. One study tried to create a map showing EV energy consumption on each road link based on link-specific traffic patterns (e.g., average speed) and geometry attributes (e.g., number of lanes, link width) [20]. Other studies looked at evaluating EV driving range in sustaining future trips [19, 21, 50, 52, 58].

Figure 2 summarizes the existing relevant studies. As seen from the figure, the number of references to energy consumption of EVs has increased since 2011, plateaued between 2015 and 2018, and then rocketed in 2019. By further differentiating the references by the source of data (i.e., real world vs. simulation), we observe that EV energy estimation models based on real-world data dominate the literature and the share of macroscale models keeps increasing, which implies a promising trend of macroscopic applications in transportation planning and operations for EVs. When differentiating the references by vehicle type, we observe that the majority of existing studies have focused on passenger cars, whereas a significant knowledge gap is present for other vehicle categories, such as transit vehicles, heavy-duty trucks, trains or locomotives, and nonroad vehicles (e.g., construction and agriculture equipment). They start to appear in the literature only after 2017, which may be due to their slower electrification process compared to passenger cars.

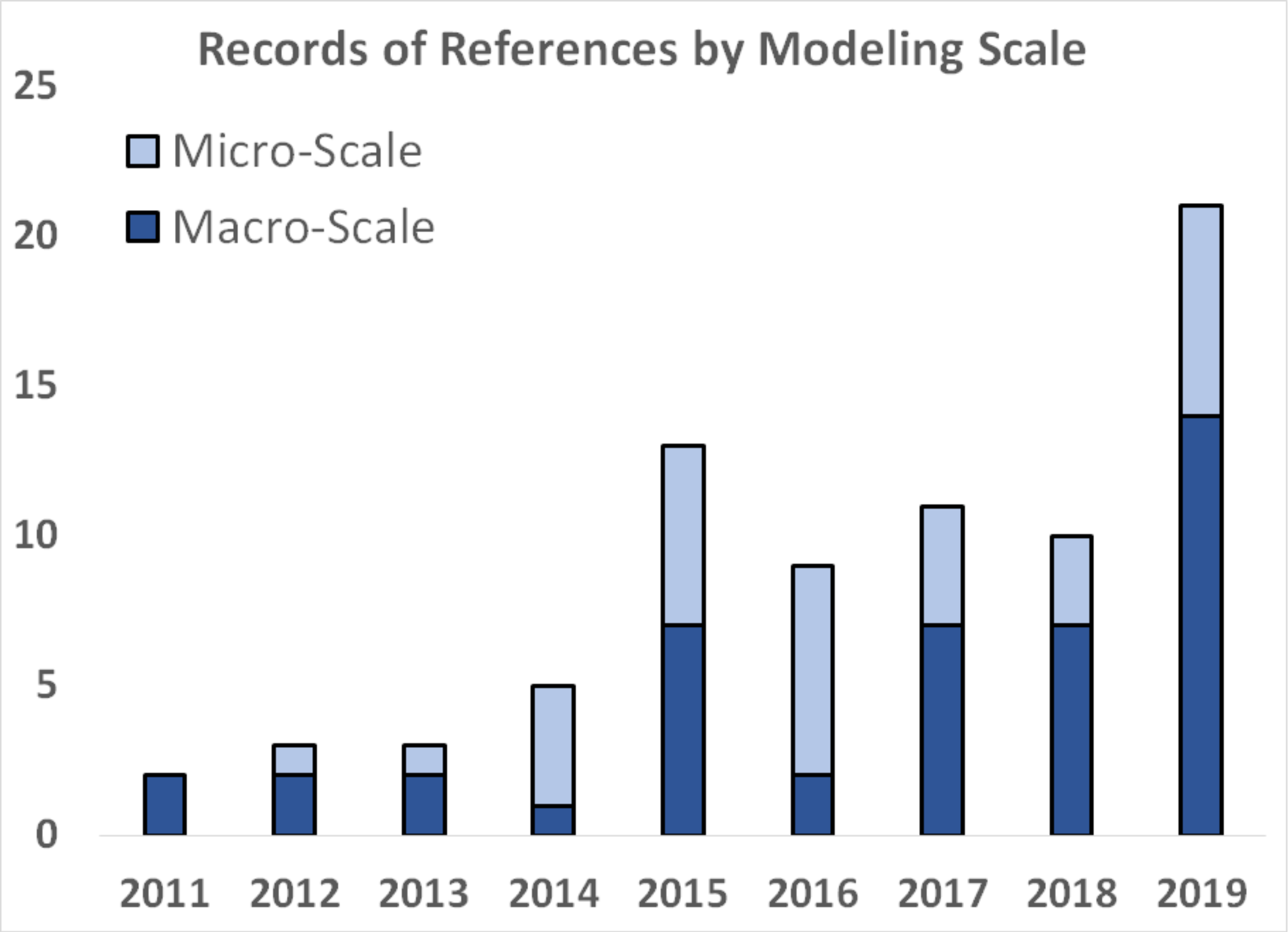

Fig. 2: Number of research papers for microscale and macroscale EV energy consumption estimation model for each calendar year.

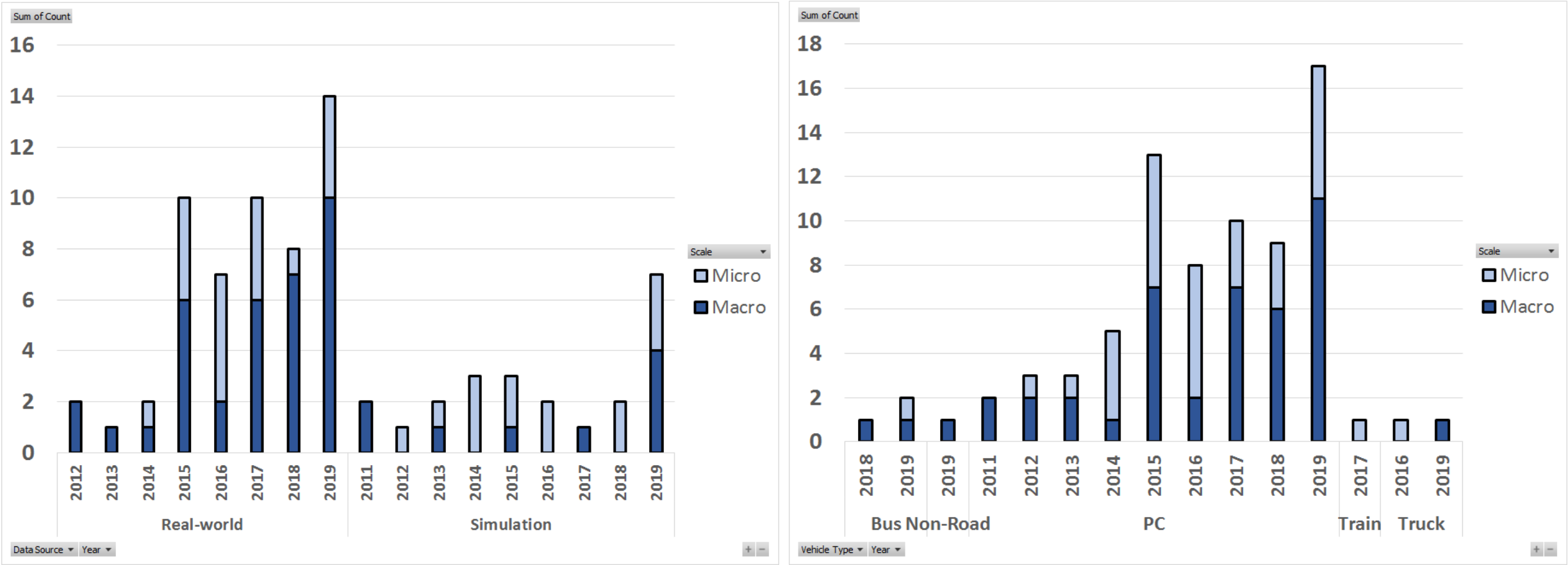

Fig. 3: Number of research papers for microscale and macroscale EV energy consumption estimation model by data source (Fig. 3a, left) and by vehicle type (Fig. 3b, right) for each calendar year.

## 5. Modeling Methodology

According to our review of the literature, existing EV energy modeling methods can be classified into three categories: rule-based, data-driven, and hybrid, as shown in Figure 4.

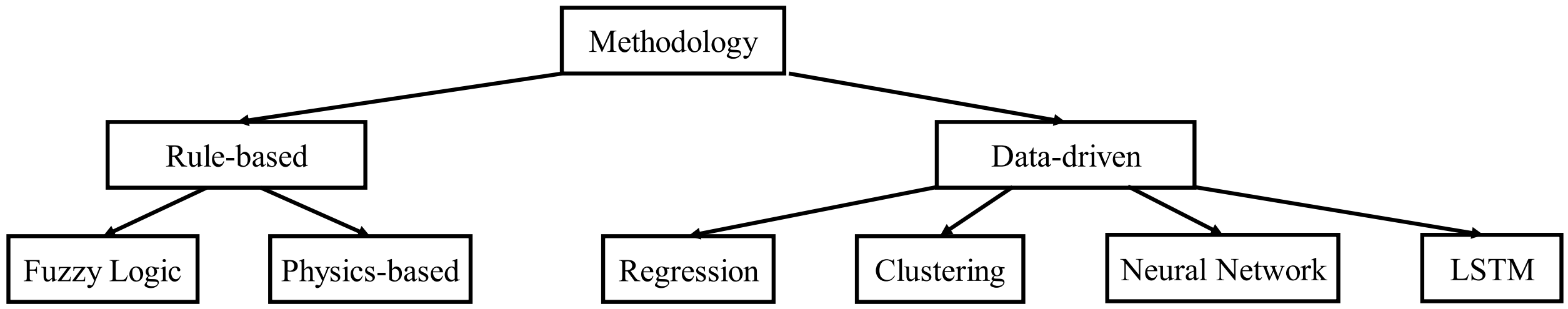

**Fig. 4: The classification scheme for the state-of-the-art EV energy modeling methods.**

### *5.1. Rule-Based*

The configuration and dynamics of battery EVs are well defined in Section 2. Compared to an internal combination engine-powered vehicle, the configuration and energy flow of an EV is less complicated. In addition, the component-wise energy efficiency of an EV is less varied. Therefore, many studies estimated the EV energy consumption rate (i.e., in the microscale) by following the Newton's law to calculate the tractive power at the wheels and assuming constant powertrain efficiency [68-72]. Others developed EV energy estimation models based on car-following models and downstream traffic information [73]. To model the regenerative braking effects, some researchers applied fuzzy logic [74], whereas others assumed a simple relationship (e.g., as a piecewise linear function) with the vehicle's speed [75]. Although rule-based models are relatively simple, their modeling accuracy when applied to a specific vehicle or scenario may not be satisfactory. In addition, there are a few challenges to extend the application of these models to a macroscopic scale: estimation errors may be accumulated if the energy consumption by a fleet of EVs is concerned, and energy impacts due to the interaction among different players are difficult to model in a complicated system (e.g., a region-wide transportation scenario).

### *5.2. Data-Driven*

Thanks to improvements in sensors, automotive electronics and telematics in recent years, more data (in terms of both type and amount) became available for EV energy consumption modeling, and a variety of data-driven techniques have been applied. To date, the most widely used statistical method is multivariate linear regression [76-80]. These EV energy consumption models usually include instantaneous speed of different orders, acceleration of different orders, and their interaction terms as the independent variables and assume their relationship with a linear predictor function. Due to rapid advancements in advanced machine learning techniques and high-performance computing, a few representative algorithms such as artificial neural networks [43] and long short-term memory [17] have been employed to estimate EV energy consumption. For example, artificial neural network models are used to construct highly nonlinear and high-order interaction terms between response factors, such as speed, acceleration, etc., and microscopic (e.g., 1Hz) and macroscopic (e.g., trip-level) energy consumption of EVs. In addition, other unsupervised learning methods (e.g., clustering, principal component analysis) have been used for data preprocessing or pattern recognition [21, 35, 55, 57]. They are widely used to classify application scenarios to improve model accuracy and select the dominant influential variables or reconstruct powerful predictors for model interpretability and generalizability.

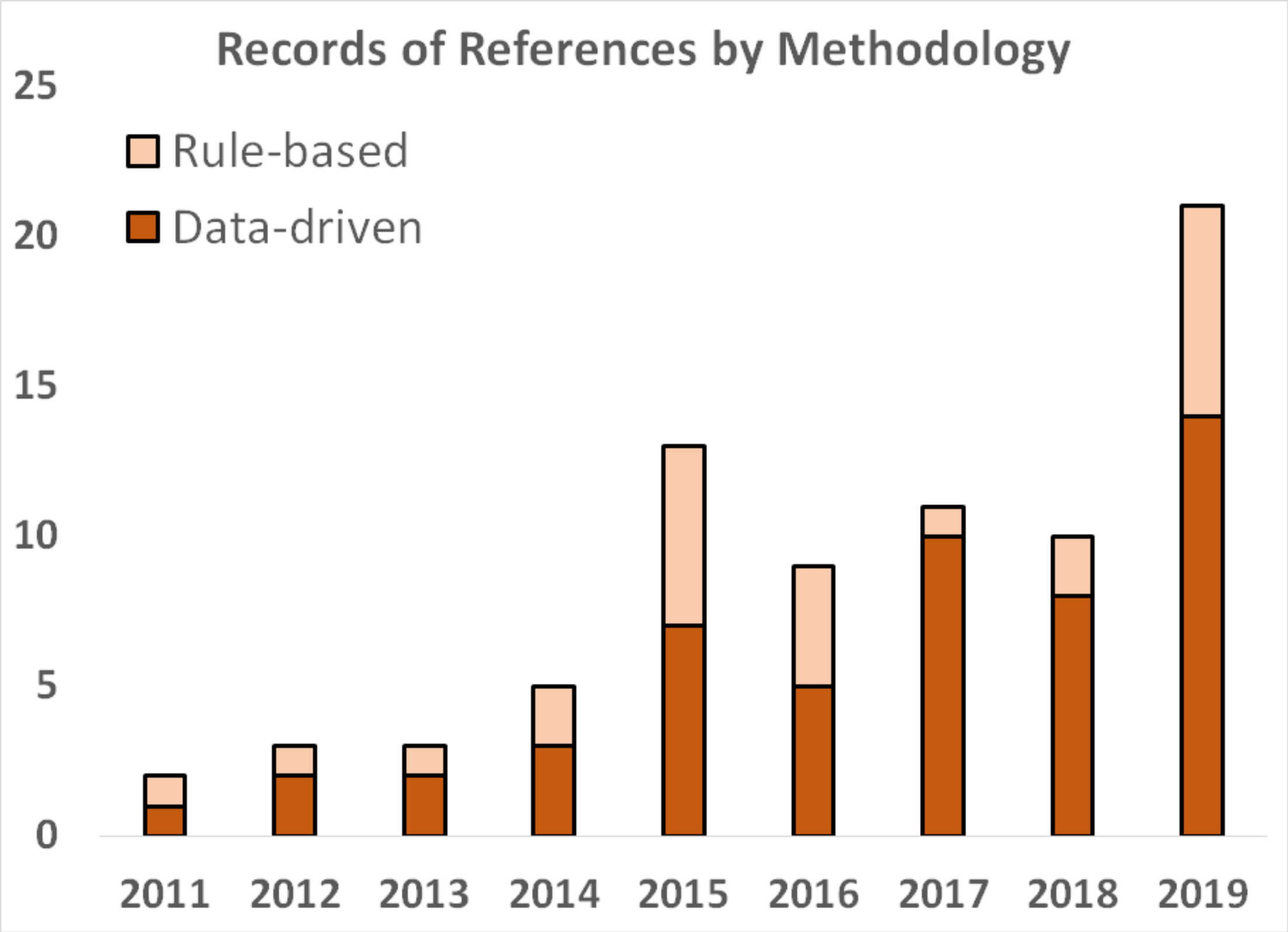

**Fig. 5: Number of research papers for EV energy consumption estimation model for each calendar year.**

Figure 5 shows that even though the volume of literature related to EV energy consumption has increased during the past few years, the number of data-driven models started to increase only after 2015. However, data-driven models became the dominant methodology in recent years. A further glimpse at the records by data source and vehicle type reveals that many studies have focused on the development of data-driven models and calibration of model coefficients with real-world data. If the vehicle type is differentiated, the data-driven methods have been dominantly used to model the energy consumption of electric passenger cars, buses, and nonroad vehicles, whereas rule-based methods are dominant in estimation models for trucks and trains.

Unlike rule-based methods that usually target microscopic models, data-driven methods can be applied to various data sources, such as powertrain and vehicle dynamics, traffic information, driving behavior, network profile, and meteorological conditions, with different spatiotemporal aggregation levels (e.g., link based, every 5 minutes). In addition, the developed models are usually customized for specific vehicles, drivers, or scenarios and their results may be very dataset dependent, making it quite challenging to generalize the conclusions.

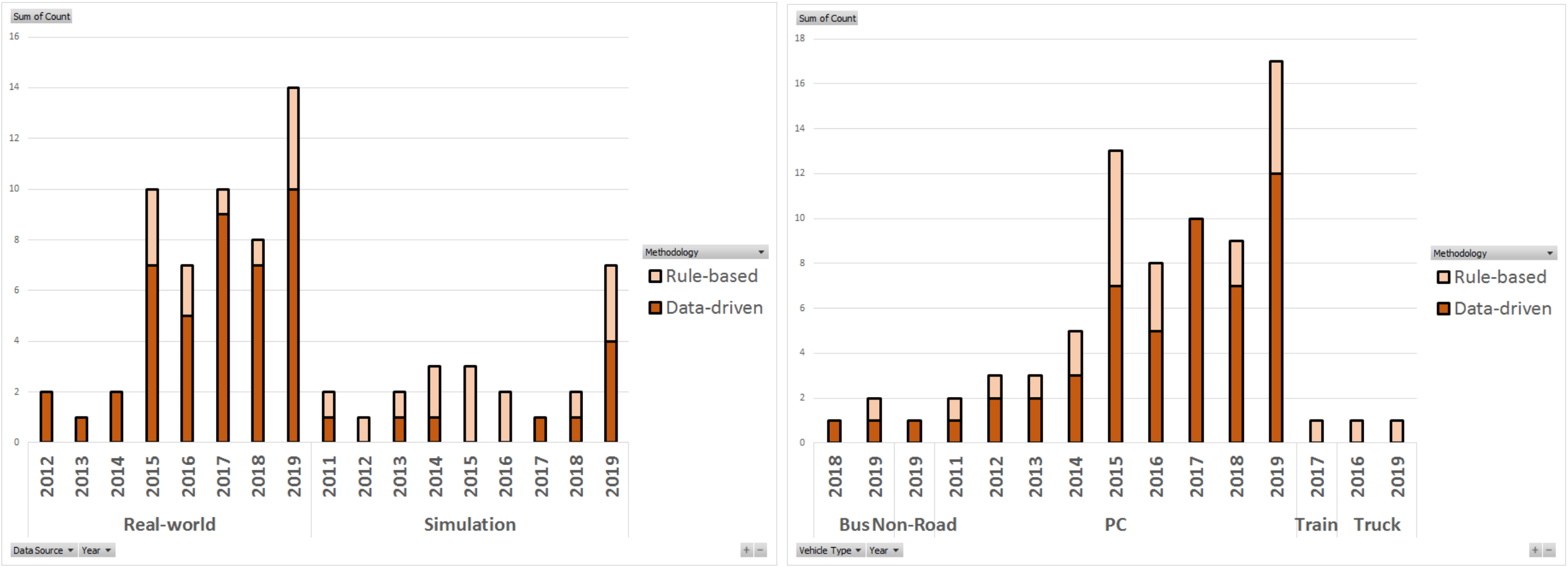

**Fig. 6: Number of research papers for EV energy consumption estimation model using rule-based and data-driven methods by data source (Fig. 6a, left) and by vehicle type (Fig 6b, right) for each calendar year.**

Despite all the advantages, data-driven EV energy consumption models have two main limitations. First, data-driven estimation models might not perform satisfactorily outside their trained or test datasets. Therefore, it is important to ensure that the datasets used for training are representative of the entire population and contain sufficient variety of information. Second, most recent data-driven models that rely on neural networks are black-box models. Although they have the potential to provide satisfactory accuracy, they are limited in explaining details of the different parameters and their implications for energy consumption.

To balance the advantages of both rule-based methods (model simplicity and generalization) and data-driven methods (model accuracy and customization), Ye et al. [24] proposed a hybrid approach to estimating the EV energy consumption rate. In this approach, the feature selection is rooted in the physical principles instead of specific datasets, whereas the model coefficients or system parameters are trained from the data to achieve customized performance for specific scenarios.

## 6. Case Study on Electric Bus Energy Consumption Estimation

We conducted a case study that utilized real-world vehicle driving and energy consumption data to estimate the energy consumption of a bus under real-world conditions. Specifically, we evaluated the impacts of including different variables and adopting different modeling methods on both microscopic and macroscopic energy consumption estimation.

The vehicle driving and energy consumption data used in this validation were collected from on-board measurement equipment on three electric buses operated in Chattanooga, Tennessee. The data collection process started in July 2018 and is an ongoing effort. The buses are BYD K9S, which are manufactured by one of the largest EV manufacturers in the world, BYD Auto Co., Ltd. The collected information included real-time GPS locations (i.e., latitude, longitude), vehicle activities (instantaneous speed, acceleration, and RPM), energy-related parameters (i.e., energy consumption rate, state of charge), and meteorological data (ambient temperature and humidity), all at 1Hz frequency. The driving cycles of transits running on those routes are shown in Figure 7. The driving cycles have a speed up to 60 to 80 kph and an acceleration range of -2 to 2 kph per second.

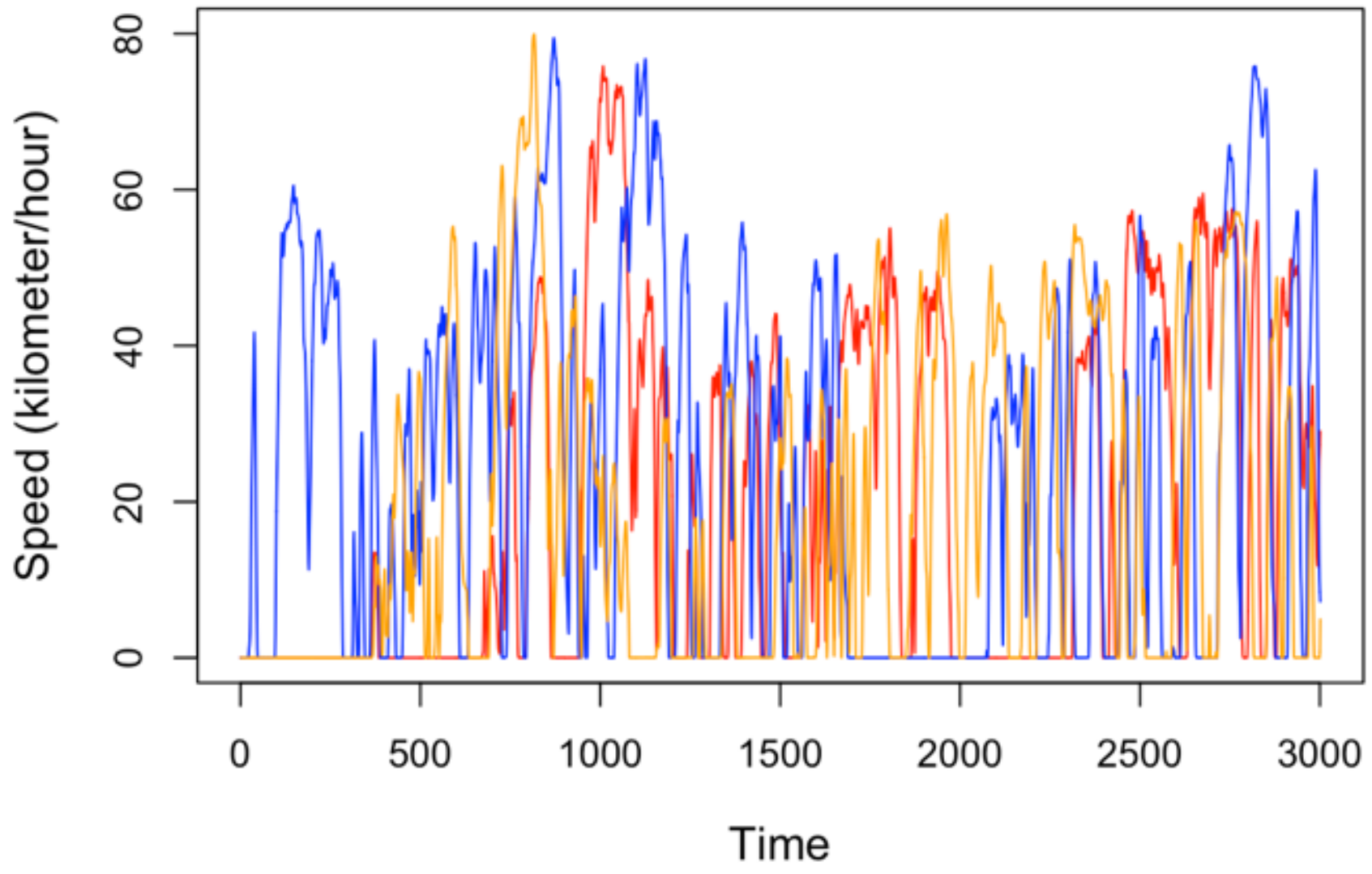

**Figure 7. Driving cycles of buses**

We summarize the input variables for both micro and macro prediction models in Table 2.

**Table 2. Description of Input Variables for Microscopic and Macroscopic Energy Consumption Model**

| | Symbols | Descriptions |
|---|---|---|
| Microscopic | $t$ | Index of second |
| | $v_t$ | Instantaneous speed at second *t* (km / h) |
| | $a_t$ | Instantaneous acceleration at second *t* (km / h / second) |
| | $d_t$ | Instantaneous motor speed at second *t* (revolution / min) |
| | $w_t$ | Ambient temperature at second *t* (Celsius degree) |
| | $q_t$ | Ambient humidity at second *t* (%) |
| | $g_t$ | Road grade at second *t* (%) |
| | $VSP_t$ | Instantaneous vehicle-specific power at second *t* (kW/ton) |
| Macroscopic | $v_l$ | Average speed of the vehicle passing through link *l* (km / h) |
| | $w_l$ | Average ambient temperature of the vehicle through link *l* (C) |
| | $q_l$ | Average humidity of a vehicle through link *l* (%) |
| | $g_l$ | Net road grade of the vehicle passing through link *l* (%) |
| | ${v_l}^2$ | Square speed of the vehicle passing through link *l* |
| | ${v_l}^3$ | Cubic speed of the vehicle passing through link *l* |

With the input variables at micro and macroscopic levels, we constructed models with different combinations of input variables and linear regression or neural network modeling methods. These two methods are most common in literature. We randomly divided the data into training and testing datasets and compare their prediction performance. We adopt mean absolute scaled error (MASE) as a measure of forecast accuracy to compare the performance of the energy consumption rate estimation model. Hyndaman and Koehlers [111] proposed MASE to serve as a generally applicable forecast error measurement that is independent of the scale of data and has better predictable behavior when the dependent variable is close to zero. MASE is calculated as follows:

$$MASE = \frac{1}{n}\sum_{j=1}^{n}\left(\frac{|P_j - \hat{P}_j|}{\frac{1}{n-1}\sum_{i=2}^{n}|P_i - \hat{P}_{i-1}|}\right)$$

where $P$ is observed energy consumption rate, $\hat{P}$ is the estimated energy consumption rate, and $n$ is the total number of observations in the testing dataset. We summarize the various model setup and their prediction performance in terms of MASE in Table 3. Model 1 to Model 3 are microscopic models containing predictive variable on dynamics (Model 1); dynamics and components (Model 2); and dynamics, components, and environment (Model 3). The prediction performance comparison shows that the MASE values of neural network models are always lower, indicating better prediction accuracy, than those of linear regression of the same model. And inclusion of more predictive variable led to better prediction performance. Similar patterns were observed in macroscopic models as well.

**Table 3. Optimal Model Configuration and Performance Metric**

| | Predicting Variable | Methodology | MASE |
|---|---|---|---|
| Microscopic | Model 1: $v_t, a_t$ | LR | 4.92 |
| | | ANN | 1.89 |
| | Model 2: $v_t, a_t, d_t, VSP_t$ | LR | 4.15 |
| | | ANN | 1.48 |
| | Model 3: $v_t, a_t, d_t, VSP_t, w_t, q_t, g_t$ | LR | 2.46 |
| | | ANN | 0.33 |
| Macroscopic | Model 4: $v_l, v_l^2, v_l^3$ | LR | 4.56 |
| | | ANN | 1.25 |
| | Model 5: $v_l, v_l^2, v_l^3, w_l, q_l, g_l$ | LR | 2.75 |
| | | ANN | 0.40 |

Note: LR = linear regression; ANN = artificial neural network

## 7. Discussion

It is self-evident that different energy consumption models are suitable for different energy-focused EV applications. Figure 7 presents some typical scenarios of EV energy model applications across scales. For example, energy consumption rate models are applicable to the development of eco-driving (mainly longitudinal maneuvers) systems for individual EVs or eco-friendly traffic signal control at intersections. Aggregated models may fit well for region-wide EV applications considering a long-term effect.

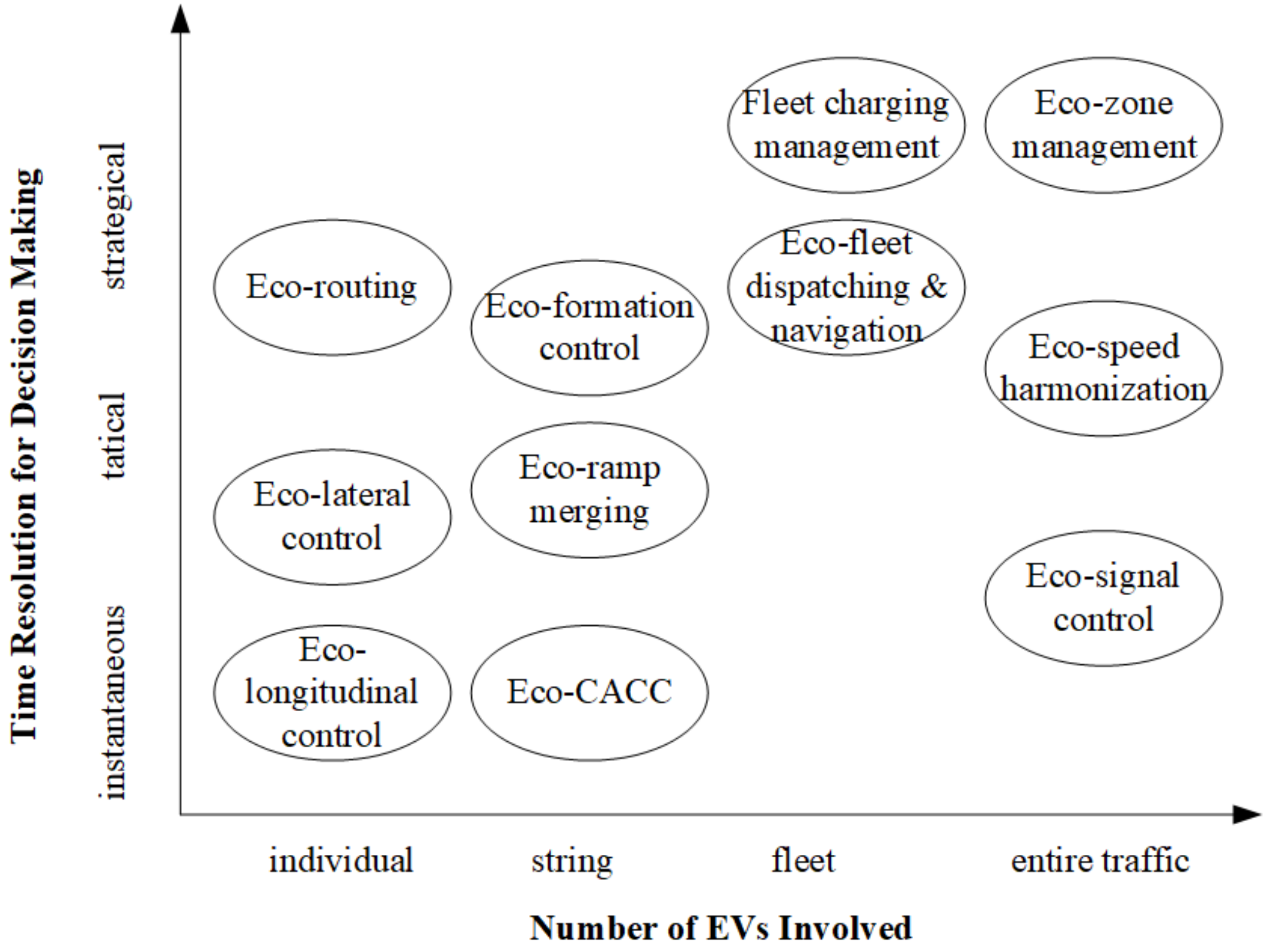

Fig. 8: EV energy model applications under different time resolutions.

The results of this review indicate some research areas that may require more attention:

1) *Energy estimation for vehicles other than passenger cars:* The majority of up-to-date literature on EV energy estimation model has focused on passenger cars. However, significant progress has been witnessed in the development of freight and transit electrification. Given the fact that trucks consume more than 20% of total transportation energy [81] and public transit is deemed a viable solution to transportation challenges in many developing countries or populated areas, more research should be conducted on modeling the energy consumption of electric trucks and buses.

2) *Application for vehicle-to-grid integration:* Vehicle-to-grid technology is considered as an emerging and cost-effective solution to optimizing both EV use and power grid operation in a cooperative manner. One challenge in vehicle-to-grid integration systems is finding a computationally efficient algorithm that can handle real-time EV energy consumption analysis and large-scale charging facility scheduling optimization [82]. Specifically, the embedded EV energy consumption model should be capable of estimating and comparing EV energy consumption over different candidate driving scenarios.

3) *Development of multiscale EV energy estimation models:* Multiscale models can simultaneously cover important features at different resolutions of time and space. Such an integrated modeling approach may preserve information at different levels, from individual components to traffic, in a collective manner. All existing studies have focused on either microscale or macroscale EV energy estimation. It will be useful to develop multiscale EV energy estimation models that can provide consistent information for energy estimation across scales. However, one major challenge would be to develop algorithms or methodologies to find accurate and efficient solutions to multiscale modeling problems.

## 8. Conclusion

This paper presents an overview of recent research efforts in the area of EV energy consumption estimation. Energy consumption estimation models were reviewed in terms of influential variables (vehicle components, driving dynamics, traffic, environment), modeling scale (microscopic vs.

macroscopic), and methodology (rule-based vs. data-driven). The properties of the data used for these models were also reviewed, including the source of data (simulation vs. real world), type of vehicles to be modeled (car, truck, bus, train, or nonroad vehicles), and publication year (2011–2019).

Vehicle component factors determine the operation of key parts for propulsion and energy flow in vehicles. They are naturally used in rule-based models at the micro scale because their changes can instantaneously and directly influence energy output of the electric motor. But certain aggregated formats of these factors are also witnessed in data-driven macroscale models in recent literature. Vehicle dynamics factors represent the motion of vehicles. The instantaneous or aggregated formats of vehicle dynamics are most widely used for EV energy modeling, regardless of the modeling scale or methodology. Traffic condition factors can be used to supplement information provided by vehicle dynamics. Traffic at one time is equal to instantaneous speed in vehicle dynamics. Therefore, microscale models do not consider traffic factors. Macroscale models use traffic factors as a proxy for vehicle dynamics in a certain period of time. Environment factors mainly relate to roadway characteristics or meteorological conditions. They are commonly used in macroscale models, especially road grade, which is used in both micro- and macroscale models.

According to the summary, data-driven EV energy consumption estimation and its applications have been attracting increasing research attention in recent years, whereas rule-based models dominate earlier literature. In addition, a growing number of macroscale models in the literature has been observed in recent years. Although models with different scales may be developed to serve different purposes and application scenarios, efforts toward multiscale model development would be valuable as an integrated solution to preserving information consistency from various spatiotemporal resolutions and aggregated levels.

**Acknowledgments:** This material is based on work supported by the Department of Energy, Office of Energy Efficiency and Renewable Energy (EERE), under Award Number DE-EE0008467.